\documentclass[floats,floatfix,showpacs,amssymb,prl,twocolumn,superscriptaddress,nofootinbib]{revtex4-1}
\usepackage[normalem]{ulem}
\usepackage{booktabs}
\usepackage{graphicx}%
\usepackage{dcolumn}%
\usepackage{bm}%
\usepackage{amsmath}
\usepackage{xcolor}
\usepackage{xspace}
\newcommand{\dl}{luminosity distance\xspace}
\newcommand{\DL}{$\mathrm{D}_\mathrm{L}$\xspace}
\newcommand{\hubbleunits}{km s$^{-1}$ Mpc$^{-1}$\xspace}
\newcommand{\rateunits}{\ensuremath{\mathrm{Gpc}^{-3} \mathrm{yr}^{-1}}\xspace}
\newcommand{\msun}{M$_\odot$\xspace}
\newcommand{\hnot}{\ensuremath{\mathrm{H}_0}\xspace}
\newcommand{\tjn}{\ensuremath{\theta_\mathrm{JN}}\xspace}
\newcommand{\degg}{\ensuremath{\mathrm{deg}^2}\xspace}
\newcommand{\secname}[1]{\textit{#1}\xspace}

\newcommand{\imr}{\texttt{IMRPhenomPv2}}

\newif\ifthetajn \thetajnfalse
\newcommand\prlsec[1]{\vspace{2mm}\noindent \emph{#1}--}

\begin{document}

\title{Measuring the Hubble constant with neutron star black hole mergers}

\author{Salvatore Vitale}
\affiliation{LIGO Laboratory and Kavli Institute for Astrophysics and Space Research, Massachusetts Institute of Technology, Cambridge, Massachusetts 02139, USA }
\email{salvatore.vitale@ligo.org}
\author{Hsin-Yu Chen}
\affiliation{Black Hole Initiative, Harvard University, Cambridge, Massachusetts 02138, USA}

\begin{abstract}
The detection of GW170817 and the identification of its host galaxy have allowed for the first standard-siren measurement of the Hubble constant, with an uncertainty of $\sim 14\%$.
As more detections of binary neutron stars with redshift measurement are made, the uncertainty will shrink. The dominating factors will be the number of joint detections and the uncertainty on the luminosity distance of each event. 
Neutron star black hole mergers are also promising sources for advanced LIGO and Virgo. If the black hole spin induces precession of the orbital plane,  the degeneracy between luminosity distance and the orbital inclination is broken,  leading to a much better distance measurement. In addition neutron star black hole sources are observable to larger distances, owing to their higher mass. 
Neutron star black holes could also emit electromagnetic radiation: depending on the black hole spin and on the mass ratio, the neutron star can be tidally disrupted resulting in electromagnetic emission.
We quantify the distance uncertainty for a wide range of black hole mass, spin and orientations and find that 
the 1-$\sigma$ statistical uncertainty can be up to a factor of $\sim 10$  better than for a non-spinning binary neutron star merger with the same signal-to-noise ratio. 
The better distance measurement, the larger gravitational-wave detectable volume, and the potentially 
bright electromagnetic emission, imply that spinning black hole neutron star binaries can be the optimal standard siren sources as long as their
astrophysical rate is larger than $O(10)$ Gpc$^{-3}$yr$^{-1}$, a value allowed by current astrophysical constraints.
\end{abstract}

\maketitle

\prlsec{Introduction}  A measurement of the local value of the Hubble parameter is crucial for our understanding of the evolution of the Universe. Over the last many years, measurements of the Hubble constant based on supernovae~\cite{2016ApJ...826...56R} or on the cosmic microwave background~\cite{2016A&A...594A..13P} got more and more precise. 
While we are now fully in the era of precision cosmology, accuracy is still elusive, with the two methods disagreeing at a $\sim 3\sigma$ level~\cite{2017NatAs...1E.169F}. 

Gravitational-wave detections can provide a totally independent way of measuring the Hubble constant, if an electromagnetic counterpart is found~\cite{SchutzNature}. 
This has been spectacularly demonstrated with the binary neutron star (BNS)  merger GW170817 and the kilonova AT 2017gfo~\cite{2017Natur.551...85A}.

In case of a positive redshift measurement with electromagnetic (EM) facilities, the uncertainty in the measurement of the Hubble constant is typically dominated by the precision with which the \dl can be measured in the gravitational-wave (GW) sector.
For example, for GW170817 one has \DL$=43.8^{+2.9}_{-6.9}$~Mpc (68.3\% confidence interval), i.e. a relative 1-$\sigma$ uncertainty of $\sim 11\%$~\citep{2017Natur.551...85A}. That corresponded to a measurement of the Hubble constant of $H_0=70^{+12.0}_{-8.0}$\hubbleunits, i.e. a relative  1-$\sigma$ uncertainty of $\sim 14\%$.
The rest of the error budget includes the uncertainty in the estimation of the peculiar velocity of the BNS host w.r.t. the Hubble flow.

One way to improve the measurement of the Hubble constant, is to build joint posteriors given many BNS mergers with host 
identification~\citep{2013arXiv1307.2638N,2016LRR....19....1A,2017arXiv171206531C}. 
Other methods have been proposed, which do not necessary rely on the identification of the host, but require dozens of sources~\citep{SchutzNature,2005ApJ...629...15H,2012PhRvL.108i1101M,2012PhRvD..86d3011D,2017PhRvD..95d3502D,2012PhRvD..85b3535T}.
The precision with which the GW \dl can be measured is usually limited by the well-known degeneracy between orbital inclination and \dl~\cite{GW150914-PARAMESTIM}, and is of the order of a few tens of percent (standard deviation) for BNSs~\cite{2011PhRvD..84j4020V,2014ApJ...784..119R,2014PhRvD..89b2002V,2016ApJ...825..116F}.

However, other potentially EM-bright type of mergers exist, for example, neutron star black hole (NSBH) mergers.
Electromagnetic~\cite{1991AcA....41..257P,1992ApJ...395L..83N,1999ApJ...527L..39J}  and neutrino~\cite{2013ApJ...776...47D,2018PhRvD..97b3009K} emission from NSBH mergers would be powered by tidal disruption of the neutron star  and the resulting accretion disk.
Whether tidal disruption happens depends on the mass ratio of the system, the spin of black hole, and the neutron star equation of state~\cite{2009PhRvD..79d4024E,2013PhRvD..87h4006F}. 
If the neutron star tidal disruption radius is larger than the innermost stable circular orbit of the system, 
the neutron star is tidally disrupted. Specifically, one would expect tidal disks when the mass ratio is smaller than a few to one and/or the black hole spin is large
~\citep{2011PhRvD..84f4018K,2012PhRvD..86l4007F,2013CQGra..30m5004L,2015MNRAS.448..541J,2015MNRAS.446..750F}. 

Black hole spins can break the degeneracy between \dl and inclination, resulting in more precise constraints on the parameters of the source~\cite{2004PhRvD..70d2001V}.
For a canonical 10-1.4\msun NSBH the \dl uncertainty can be a factor of few smaller than what achievable with a typical BNS~\citep{2014PhRvL.112y1101V} .

There are two main reasons why. On one side, the degeneracy between \dl and inclination is only present in the inspiral phase of the GW signal, whereas accessing the merger and ringdown can help resolving it~\citep{ 2011PhRvD..84f4003M,2016PhRvD..93b4003K}. As the merger frequency decreases for increasing total mass, the \dl of NSBH sources can be measured better than that of BNS coalescences. 
Additionally, NSBHs can have significant spin precession, as long as the black hole spin is not negligible and is not aligned with the orbital angular momentum.
Spin precession gives the waveform a characteristic phase and amplitude modulation~\citep{PhysRevD.49.6274}, which significantly reduces the degeneracy with the inclination angle~\cite{2014PhRvL.112y1101V,Vitale:2016avz}.
However, NSBHs are expected to merge less often~\citep{2016ApJ...832L..21A}. In fact, no NSBH has been discovered to date, either in the EM or in the GW band. 
Non-detection of NSBHs during LIGO's first science run allowed estimation of their merger rate to be smaller than $3600$~Gpc$^{-3}$yr$^{-1}$~\citep{2016ApJ...832L..21A}.

In this letter we show that NSBHs can potentially serve as a competitive standard sirens to BNSs, if their merger rate is larger than $O(10)$ Gpc$^{-3}$yr$^{-1}$, a value allowed by current constraints.

\prlsec{Method} 
We simulate  NSBH and BNS sources and add them 
into ``zero-noise'' (which yields the same results that would be obtained averaging over many noise realizations~\citep{2008PhRvD..77d2001V,2014ApJ...784..119R}). We work with a network made by the two LIGO detectors and the Virgo detector at their design sensitivity~\citep{Aasi:2013wya}. This choice does not significantly affect our results, since we are mostly interested in the \emph{ratio} of uncertainties for NSBHs and BNSs, which is not a strong function of the exact sensitivity curve.

While it is likely that the EM brightness of NSBHs depends on the mass ratio, spin magnitude and spin tilt angle, the exact dependence is not known.
We therefore do not restrict our analysis to a particular combination of mass and spins, but rather cover a large 
range of possibilities. We consider three different NSBH masses: $10-1.4$\msun, $7.5-1.4$\msun and $5-1.4$\msun. We do not  assign spin to the neutron star (consistent with the fact that known neutron stars have very small spins). If neutron stars turn out to be significantly spinning, that would actually improve the measurement of \dl, by adding extra precession.
For each system, we consider three possible orientations of the black hole spins. We will refer to the angle between the black hole spin and the orbital angular momentum as the tilt angle. In case of precession, both the spin vector and the orbital angular momentum precess around the total angular momentum (whose direction is nearly fixed in space~\citep{2017PhRvD..96b4007Z}). We quote tilt angles at a reference frequency of $20$~Hz, corresponding to our choice for the lower frequency of the gravitational-wave analysis. The three values we use are $\tau_{\mathrm{BH}}=(0^\circ, 60^\circ, 90^\circ)$.  
A tilt angle of $0^\circ$ means that the spin vector is aligned with the orbital angular momentum. In this case no precession happens, and one would expect the degeneracy between distance and inclination to still be present. Conversely, $90^\circ$ implies maximum precession. The results we obtain for those two extreme cases will thus bracket what one can expect.
For each of the tilt angles, we consider two possible values of the dimensionless BH spin magnitude~\citep{GW150914-PARAMESTIM}: moderate ($0.5$) and large ($0.89$).

Both BNS and NSBH signals are generated using the \imr{} waveform family~\citep{2015PhRvD..91b4043S,Hannam:2013oca} and have a network signal-to-noise ratio (SNR) of 20. 
While GW170817 was much louder, with a network SNR of 32.4~\cite{2017PhRvL.119p1101A}, an SNR of 20 is more representative of what will be typical. We put all sources at the same sky position, near the maximum of LIGO's antenna patterns, where we would expect the typical detection to be made~\cite{2017ApJ...835...31C}.
To check that the results we obtain are solid, we also considered a second sky position (near the north pole direction) and verified the main conclusions are the same. In what follows, we will thus only show plots obtained with sources near the maximum of LIGO's antenna patterns.

The effects of spin precession in the detector frame are not only dependent on the actual degree of precession, 
but also on the inclination angle, defined as the angle between the line of sight vector and the 
\emph{total} angular momentum, $\theta_{\mathrm JN}$~\cite{PhysRevD.49.6274}. The effects of precession are more 
visible if the system is observed at inclinations close to $90^{\circ}$ (edge-on)~\cite{2014PhRvL.112y1101V,Vitale:2016avz}.

To capture that dependence, we repeat all simulations at several values of inclination angle, uniformly spaced in $\cos\theta_{\mathrm{JN}}$, while keeping the SNR fixed to 20.
%. Critically, the \emph{SNR} is always kept at the same value of 20. Therefore, when the inclination angle of a simulated source is varied, its true \dl is also changed to yield the same SNR.
%This implies that all observable differences are uniquely due to the different morphology of the signals.

%
We use the \texttt{LALInference} sampler~\cite{2015PhRvD..91d2003V} to estimate the parameters of the sources, and the Reduced Order Quadrature (ROQ) approximation to the likelihood~\cite{2015PhRvL.114g1104C} to speed up the computation. We notice that the ROQ method has only been tuned up to spin magnitude of 0.89~\cite{Smith:2016qas}, which explains our choice for the maximum spin.
We also stress that the \imr{} waveform family does not contain higher order harmonics, which might play a role for large mass ratios. This choice is forced on us by the lack of fast-to-compute inspiral-merger-ringdown waveforms with precessing spins. While we do not expect the results to significantly change, this study should be repeated once more sophisticated waveform models are available.
Similarly, we assume that the compact objects are in quasi-circular orbits, i.e. we neglect eventual eccentricity. This is a reasonable assumption since eccentricity is expected to be radiated away very quickly, circularizing the binary's orbit~\cite{2016ApJ...818L..22A}.

For the NSBH analysis we use the same prior shapes described by the LIGO and Virgo collaborations in Ref.~\cite{GW150914-PARAMESTIM}. The prior ranges on the component masses are given by Ref.~\cite{2015PhRvL.114g1104C}.
For the BNS events, we assume that the neutron stars have no spins, which significantly reduces the computational time. Similarly, we do not account the tidal deformability of the NS. Neither of this effects is expected to impact the measurability of the luminosity distance.
In all analyses, we assume the sky position of the source is known, since we work under the assumption that an electromagnetic counterpart to the GW event is found, which provides the necessary redshift measurement.
We marginalize over instrumental calibration errors as in Ref.~\cite{2017PhRvL.119p1101A} assuming gaussian priors on the calibration spline points with standard deviations of 3\% for the amplitude and 1.5 degrees in phase, for all instruments. These are realistic estimates of what can be achieved by advanced detectors.

\prlsec{Results}
\begin{figure}
\centering
\includegraphics[width=1.1\columnwidth]{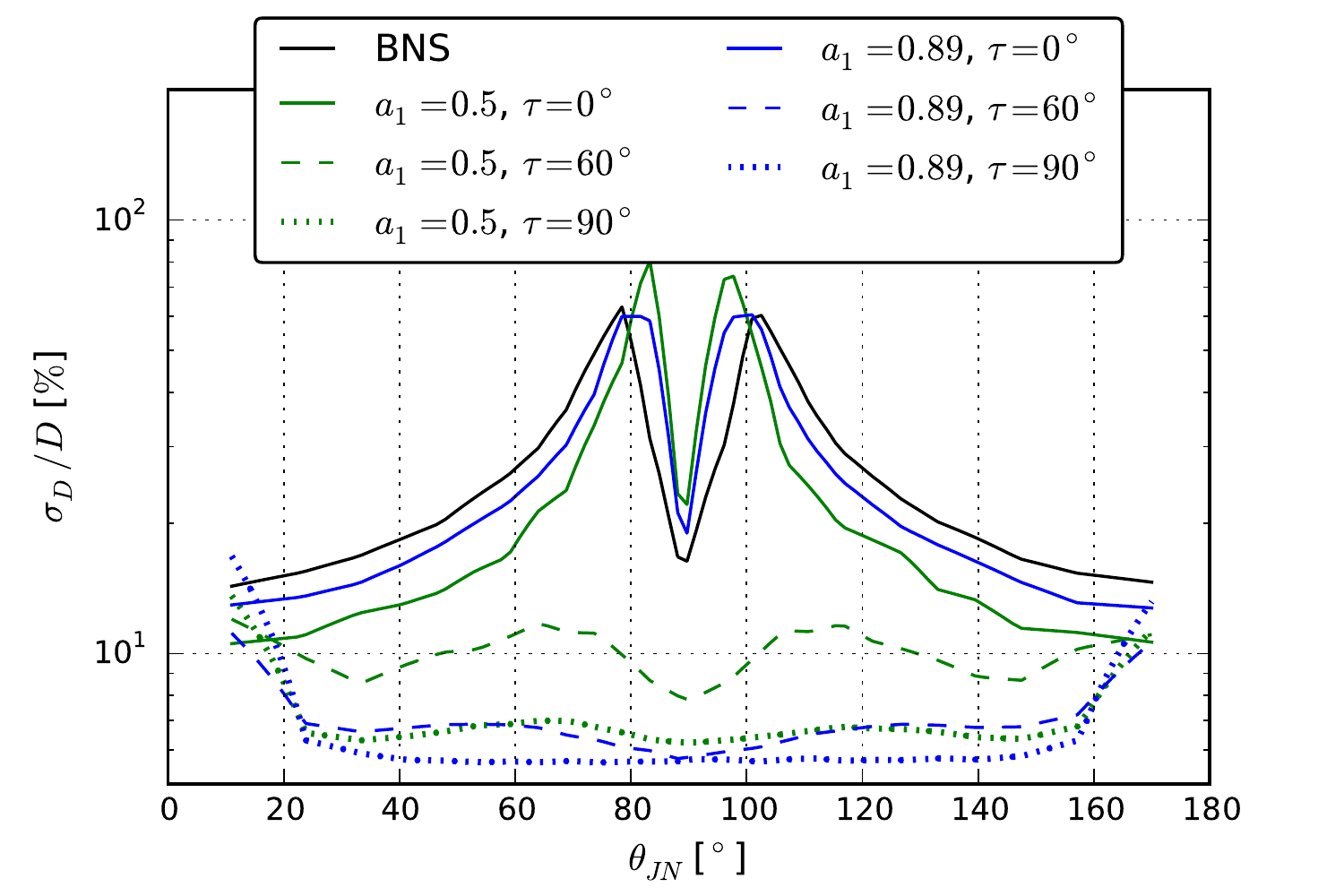}
\caption{\label{fig:fracdl}
1-$\sigma$ fractional distance uncertainty (in percent) as a function of the true inclination angle (in degrees).
}
\end{figure}
In Figure~\ref{fig:fracdl} we plot the fractional 1-$\sigma$ \dl uncertainty relative to the true distance against the true value of the inclination angle  for all the systems we simulated. 
The color allows to distinguish the BNSs (black) from the NSBHs with $0.5$ spin magnitude (green) or $0.89$ spin magnitude (blue).

Let us start by analyzing non-precessing systems (solid lines). Figure~\ref{fig:fracdl}  shows that the uncertainty steadily increases from face-on to inclinations quite close to edge-on. 
It is important to remember that this is the \emph{relative} uncertainty. Since all the sources are kept at the same SNR, 
binaries at higher inclination angles have to be \emph{closer} to yield an SNR of 20.  This is why the uncertainty in 
Figure~\ref{fig:fracdl} goes up for non-precessing systems. The actual uncertainty is roughly constant, but the true \dl gets smaller and smaller. For the BNS systems, the 1-$\sigma$ uncertainty is roughly $35$~Mpc for inclinations in the range $[0,60]^\circ$.
When the true inclination is close to edge-on, both \dl and inclination measurement get better (that is because the cross polarization of GW goes to zero when the system is edge-on, which breaks the degeneracy with the inclination~\cite{2014PhRvL.112y1101V}), and both true and relative distance uncertainty reach a minimum.

In case of precession (dashed and dotted lines), the relative 1-$\sigma$ uncertainty can be a factor of $\gtrsim 2$ smaller than what achievable with a BNS at the same position.
The smallest uncertainties are obtained for the largest spins and tilt we considered (blue dotted). That is unsurprising: a large and misaligned BH spins results in a significant waveform amplitude modulation, which entirely breaks the degeneracy.

Similar conclusions apply to the measurement of the inclination angle itself, which could provide precious information to study the EM emission~\citep{2018Natur.554..207M,2017arXiv171203237L}. 
We find that for $\tjn\sim 30^\circ$ the 1-$\sigma$ uncertainty is  $\sigma_{\tjn}\sim 15^\circ$ in absence of precession, whereas precessing NSBHs can yield uncertainties as small as $\sigma_{\tjn}\sim3^\circ$. The difference is even larger for orientations closer to edge-on.

\begin{figure}
\centering
\includegraphics[width=1\columnwidth]{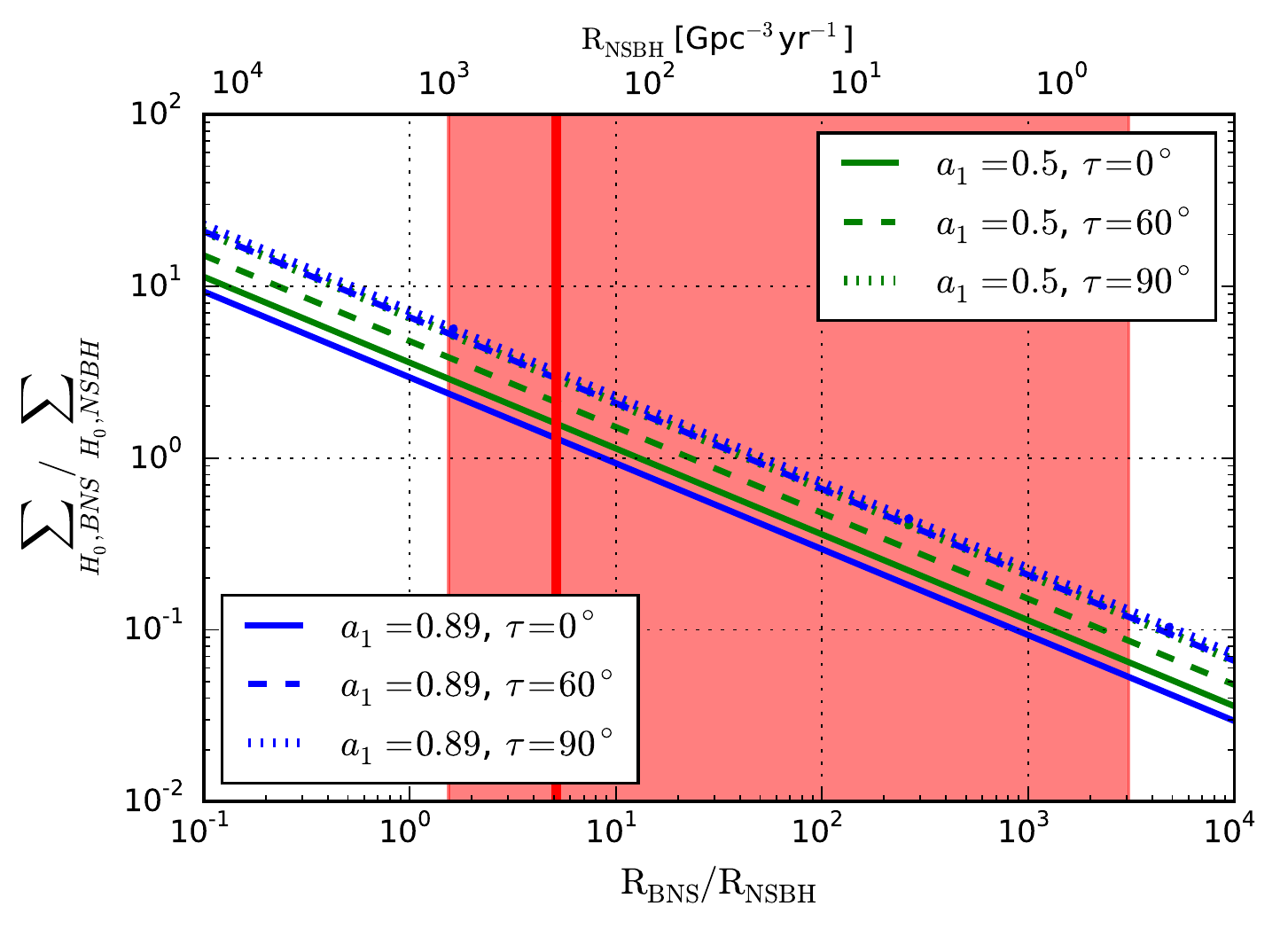}
\caption{\label{fig:h0ratio}
Relative $H_0$ 1-$\sigma$ uncertainty as a function of NSBH and BNS astrophysical rate ratio. 
See the body for more details.}
\end{figure}

In this work we assume that the Hubble velocity is perfectly measured from the redshift, in which case the uncertainty in the measurement of the luminosity distance can directly converted to the same relative uncertainty in the measurement of the Hubble constant (In practice, the Hubble velocity measurement is affected by uncertainties due to redshift calibration and peculiar velocity of the host galaxy. We will go back to this point in Section \secname{Discussion}.
The \hnot uncertainty after combining N detections can be written as $\sum_{\hnot}=\bar{\sigma}_{\hnot}/\sqrt{N}$,
where $\bar{\sigma}_{\hnot}$ is the expected \hnot uncertainty for a single event, which is numerically the same as the distance uncertainty in Section \secname{Method}. This means we are taking sources of SNR 20 as representative
While in reality sources with different SNRs will contribute to the measurement, our approach is appropriate to assess the relative precision achievable with NSBHs and BNSs.

As we have shown in Fig.~\ref{fig:fracdl}, the uncertainty in the \dl depends significantly on the inclination angle of the source, which cannot be directly averaged out, since GWs from face-on binaries are easier to detect than for edge-on binaries~\cite{Maggiore}.
Once one folds in this selection effect, the resulting distribution for the inclination angle of \emph{detectable} sources can be shown to follow a bimodal curve, with maxima at $\sim30^\circ$ and $\sim 150^\circ$ and a local minima at $90^\circ$. An analytical form for the expected distribution, which we use to weight events based on their probability of detection, is provided elsewhere~\cite{2011CQGra..28l5023S}.

We can now check if and to which extent NSBHs can contribute significantly to the measurement of \hnot. 
The answer will obviously depend on the number of NSBH detections, which in turns depends on the  (poorly known) astrophysical merger rates $R$ of NSBHs~\cite{2016ApJ...832L..21A}. 

More specifically, the number of detections for each class of source can be written as  $N=R\times V \times T$, where $R$ is the astrophysical rate, $V$ is the redshifted volume~\cite{2017arXiv170908079C}, and $T$ is the observing time (factoring the duty cycle of the detectors). 

We can thus write the \hnot uncertainty after combining all BNS detections made in the time period $T$, and compare it to what doable with the NSBHs detected in the same time:
\begin{equation*}\label{eq:rateratio}
\frac{\sum_{\hnot,{\rm BNS}}}{\sum_{\hnot,{\rm NSBH}}}={\frac{\bar{\sigma}_{\hnot,{\rm BNS}}}{\bar{\sigma}_{\hnot,{\rm NSBH}}}}\sqrt{\frac{R_{\rm NSBH}\times V_{\rm NSBH}}{R_{\rm BNS}\times V_{\rm BNS}}}
\end{equation*}

Where the observing time $T$ cancels out. For both NSBHs and BNSs, we can calculate the redshifted volume using method described in~\cite{2017arXiv170908079C}. 
We can now plot the ratio of \hnot uncertainty achievable with BNSs and NSBHs as a function of the relative astrophysical merger rate.

This is shown for the 10-1.4 \msun NSBHs in Fig.~\ref{fig:h0ratio}. The different diagonal lines refers to various values of BH spin magnitude and orientation we have considered.

For example, if the merger rates of NSBHs with BH with spin magnitude 0.5 and $60^{\circ}$ tilt are more than 1/25 of BNSs astrophysical 
rate, then NSBHs alone would yield a better \hnot constraint than what doable with BNSs.
If the NSBH population happens to have larger spins, or tilts, or both, fewer NSBH are required to achieve a 
precision comparable to BNS. In the best case, even if there is a single NSBH merger for every 50 BNSs is enough.
Conversely, in absence of spin precession the \dl estimate of each NSBH source is only marginally better 
than for BNS, and an higher relative ratio is required to achieve equal precision. In this case the actual value of 
the spin magnitude is not very important, and for both the 0.5 and the 0.89 spin magnitude we obtain that more than 1 NSBH for every 10 BNSs should merge to yield the same \hnot precision.

The vertical shaded area in Fig.~\ref{fig:h0ratio} represents a possible range of relative merger rates. Those are obtained by taking the minimum (0.5~\rateunits) , median (300~\rateunits, vertical thick line) and maximum (1000~\rateunits)  NSBH rates from Ref.~\cite{2016ApJ...832L..21A} and the median BNS rate measured after the discovery of GW170817 (1540~\rateunits)~\cite{2017PhRvL.119p1101A}. The ticks on the upper x axis give the NSBH rate assuming the median BNS rate.

The uncertainties are large for both class of sources, thus these lines should only be taken as an indication of what is possible.
In particular, we see relative rates higher than 1 NSBH per 10 BNSs are not excluded. Those rates would imply that \hnot measurement with NSBHs only (no matter of their spin) is better than what doable with BNSs.

Lower mass NSBHs would require higher rates to achieve equal uncertainty.  For example, for the $5-1.4$~\msun sources, at least 1 NSBH for every 20 BNSs is required, independently on the spin magnitude and orientation.

\prlsec{Discussion} The main result we found is that inference of \hnot with NSBHs can be better than with BNS systems, as long as the relative merger rate of NSBHs and BNSs is larger than 1/10 if all NSBHs have aligned spins or 1/50 if significant spin precession is present. Both these values are still allowed by current estimate of the merger rates of BNSs and NSBHs (Fig.~\ref{fig:h0ratio}).
In what follows we list a few caveats and possible developments of this analysis. 

The results presented in Fig.~\ref{fig:h0ratio} assume that for all detectable NSBHs an electromagnetic counterpart can be found, and hence that the probability of finding a counterpart does not significantly depend on the orbital orientation.
In reality, since EM emission in NSBHs is expected to be produced by equatorial tidal disks~\cite{2015PhRvD..92d4028K}, the probability of detecting the EM counterpart could strongly depend on the orientation angle. 
As models are made available to calculate how the EM detectability depends on the inclination angle and spins, they can be folded in while weighting how systems at different inclination angles contribute to the \hnot measurement.

While theoretical~\cite{2010PhRvD..81f4026F,2009CQGra..26l5004F,PhysRevD.87.084053} and numerical~\cite{2013PhRvD..87h4006F,2013CQGra..30m5004L,2015PhRvD..92h4050P,2018MNRAS.474L..81L} work exists, the EM emission from NSBHs is not yet fully understood. There exist models suggesting that large spin tilts can reduce the amount of ejecta~\cite{2016ApJ...825...52K}. This, of course, might reduce the fraction of NSBH sources that can contribute to the \hnot measurement. On the other hand, larger BH spin can lead to more massive accretion disks, and hence brighter EM emission~\cite{2009PhRvD..79d4024E,2013CQGra..30m5004L}.
We tried to capture some of these possible scenarios by providing results for different values of spin magnitude and orientation.

One might expect that NSBH hosts are hard to localize given their smaller bandwidth~\cite{2011CQGra..28j5021F}. 
While it is true that NSBHs will typically be localized to larger areas than BNSs, the main issue is whether the localization is so poor that the area cannot be covered by optical facilities. 
This will be less of a concern as the network of gravitational-wave detector expands with the inclusion of KAGRA in Japan~\cite{PhysRevD.88.043007} and LIGO India~\cite{M1100296}. 
Even the LIGO-Virgo network can detect heavy sources to within a few tens of square degrees (see e.g. Fig.~5 of Ref.~\cite{2017MNRAS.466L..78V}) or smaller ~\cite{2016arXiv161201471C}.
This was shown by the binary black hole GW170814, which was localized within an area of $60$~\degg, despite being sub-threshold in Virgo (SNR of $4.8$~\cite{2017PhRvL.119n1101A}). This is an area that can be comfortably covered by present (and future) optical facilities.

Another possible concern are waveform systematics. In particular, the model we used does not include tidal effects, which are important in the last stages of the orbital evolution. In general, waveform modeling for NSBH sources is extremely complex. However, the main result that can make NSBH competitive standard sirens is that their luminosity distance can be estimated precisely, due to spin precession. We expect this result to hold true even as more sophisticated waveform models are developed. This is because spin precession plays a major role at low frequency, when the orbital separation is large. Whereas the waveform models might improve at frequencies above a few hundred Hertz, at lower frequencies the current models are sufficient.
Using a very different waveform family, which does not even model merger and ringdown, one can find  improvements in the measurement of the \dl similar to what we present here~\cite{2014PhRvL.112y1101V}.

As mentioned above, a non negligible fraction of the total \hnot uncertainty for GW170817 came from the uncertainty on the peculiar velocity of the host galaxy relative to the Hubble flow.
This will be less of a problem for NSBH sirens. The average redshift of $10-1.4$~\msun NSBH detected 
by advanced detectors at design sensitivity is $\sim 0.1$, where the velocity of the Hubble flow is much larger. At that redshift a representative peculiar velocity uncertainty of $200$~km/s would contribute to $\sim 0.7\%$ of the \hnot uncertainty, which is significantly smaller than the uncertainty arising from the GW analysis. 
On the other hand, a Milky Way-like galaxy would have an apparent magnitude of 17.5 at this distance, well within reach of many EM facilities. This would allow for a systematic follow-up of the host galaxy, if the EM counterpart is identified.
In conclusion, while significant uncertainties still exist on the actual merger rate of NSBH, and on their EM emission, NSBH have the potential to significantly contribute to the measurement of the Hubble constant. 
More numerical and theoretical work on the merger and the resulting electromagnetic and neutrino emission would maximize the scientific impact of future NSBH detections.

\prlsec{acknowledgments} We acknowledge valuable discussions with Jolien Creighton, Thomas Dent, Daniel Holz, Scott Hughes, Brian Metzger, Francesco Pannarale, Bernard Schutz, Nicholas Stone, Licia Verde and John Veitch. We would like to thank the anonymous referees for their suggestions. 
SV acknowledges the support of the National Science Foundation and the LIGO Laboratory. LIGO was constructed by the California Institute of Technology and Massachusetts Institute of Technology with funding from the National Science Foundation and operates under cooperative agreement PHY-0757058.
HYC was supported by the Black Hole Initiative at Harvard University, through a grant from the John Templeton Foundation.
The authors would like to acknowledge the LIGO Data Grid clusters, without which the simulations could not have been performed. We are grateful for computational resources provided by Cardiff University, and funded by an STFC grant supporting UK Involvement in the Operation of Advanced
LIGO.
This is LIGO Document P1800094
\bibliography{references}

\begin{thebibliography}{66}
\expandafter\ifx\csname natexlab\endcsname\relax\def\natexlab#1{#1}\fi
\expandafter\ifx\csname bibnamefont\endcsname\relax
  \def\bibnamefont#1{#1}\fi
\expandafter\ifx\csname bibfnamefont\endcsname\relax
  \def\bibfnamefont#1{#1}\fi
\expandafter\ifx\csname citenamefont\endcsname\relax
  \def\citenamefont#1{#1}\fi
\expandafter\ifx\csname url\endcsname\relax
  \def\url#1{\texttt{#1}}\fi
\expandafter\ifx\csname urlprefix\endcsname\relax\def\urlprefix{URL }\fi
\providecommand{\bibinfo}[2]{#2}
\providecommand{\eprint}[2][]{\url{#2}}

\bibitem[{\citenamefont{{Riess} et~al.}(2016)\citenamefont{{Riess}, {Macri},
  {Hoffmann}, {Scolnic}, {Casertano}, {Filippenko}, {Tucker}, {Reid}, {Jones},
  {Silverman} et~al.}}]{2016ApJ...826...56R}
\bibinfo{author}{\bibfnamefont{A.~G.} \bibnamefont{{Riess}}},
  \bibinfo{author}{\bibfnamefont{L.~M.} \bibnamefont{{Macri}}},
  \bibinfo{author}{\bibfnamefont{S.~L.} \bibnamefont{{Hoffmann}}},
  \bibinfo{author}{\bibfnamefont{D.}~\bibnamefont{{Scolnic}}},
  \bibinfo{author}{\bibfnamefont{S.}~\bibnamefont{{Casertano}}},
  \bibinfo{author}{\bibfnamefont{A.~V.} \bibnamefont{{Filippenko}}},
  \bibinfo{author}{\bibfnamefont{B.~E.} \bibnamefont{{Tucker}}},
  \bibinfo{author}{\bibfnamefont{M.~J.} \bibnamefont{{Reid}}},
  \bibinfo{author}{\bibfnamefont{D.~O.} \bibnamefont{{Jones}}},
  \bibinfo{author}{\bibfnamefont{J.~M.} \bibnamefont{{Silverman}}},
  \bibnamefont{et~al.}, \bibinfo{journal}{\apj} \textbf{\bibinfo{volume}{826}},
  \bibinfo{eid}{56} (\bibinfo{year}{2016}), \eprint{1604.01424}.

\bibitem[{\citenamefont{{Planck Collaboration}
  et~al.}(2016)\citenamefont{{Planck Collaboration}, {Ade}, {Aghanim},
  {Arnaud}, {Ashdown}, {Aumont}, {Baccigalupi}, {Banday}, {Barreiro},
  {Bartlett} et~al.}}]{2016A&A...594A..13P}
\bibinfo{author}{\bibnamefont{{Planck Collaboration}}},
  \bibinfo{author}{\bibfnamefont{P.~A.~R.} \bibnamefont{{Ade}}},
  \bibinfo{author}{\bibfnamefont{N.}~\bibnamefont{{Aghanim}}},
  \bibinfo{author}{\bibfnamefont{M.}~\bibnamefont{{Arnaud}}},
  \bibinfo{author}{\bibfnamefont{M.}~\bibnamefont{{Ashdown}}},
  \bibinfo{author}{\bibfnamefont{J.}~\bibnamefont{{Aumont}}},
  \bibinfo{author}{\bibfnamefont{C.}~\bibnamefont{{Baccigalupi}}},
  \bibinfo{author}{\bibfnamefont{A.~J.} \bibnamefont{{Banday}}},
  \bibinfo{author}{\bibfnamefont{R.~B.} \bibnamefont{{Barreiro}}},
  \bibinfo{author}{\bibfnamefont{J.~G.} \bibnamefont{{Bartlett}}},
  \bibnamefont{et~al.}, \bibinfo{journal}{Astronomy \& Astrophysics}
  \textbf{\bibinfo{volume}{594}}, \bibinfo{eid}{A13} (\bibinfo{year}{2016}),
  \eprint{1502.01589}.

\bibitem[{\citenamefont{{Freedman}}(2017)}]{2017NatAs...1E.169F}
\bibinfo{author}{\bibfnamefont{W.~L.} \bibnamefont{{Freedman}}},
  \bibinfo{journal}{Nature Astronomy} \textbf{\bibinfo{volume}{1}},
  \bibinfo{eid}{0169} (\bibinfo{year}{2017}), \eprint{1706.02739}.

\bibitem[{\citenamefont{Schutz}(1986)}]{SchutzNature}
\bibinfo{author}{\bibfnamefont{B.~F.} \bibnamefont{Schutz}},
  \bibinfo{journal}{Nature} \textbf{\bibinfo{volume}{323}}, \bibinfo{pages}{310
  EP } (\bibinfo{year}{1986}),
  \urlprefix\url{http://dx.doi.org/10.1038/323310a0}.

\bibitem[{\citenamefont{{Abbott}
  et~al.}(2017{\natexlab{a}})\citenamefont{{Abbott}, {Abbott}, {Abbott},
  {Acernese}, {Ackley}, {Adams}, {Adams}, {Addesso}, {Adhikari}, {Adya}
  et~al.}}]{2017Natur.551...85A}
\bibinfo{author}{\bibfnamefont{B.~P.} \bibnamefont{{Abbott}}},
  \bibinfo{author}{\bibfnamefont{R.}~\bibnamefont{{Abbott}}},
  \bibinfo{author}{\bibfnamefont{T.~D.} \bibnamefont{{Abbott}}},
  \bibinfo{author}{\bibfnamefont{F.}~\bibnamefont{{Acernese}}},
  \bibinfo{author}{\bibfnamefont{K.}~\bibnamefont{{Ackley}}},
  \bibinfo{author}{\bibfnamefont{C.}~\bibnamefont{{Adams}}},
  \bibinfo{author}{\bibfnamefont{T.}~\bibnamefont{{Adams}}},
  \bibinfo{author}{\bibfnamefont{P.}~\bibnamefont{{Addesso}}},
  \bibinfo{author}{\bibfnamefont{R.~X.} \bibnamefont{{Adhikari}}},
  \bibinfo{author}{\bibfnamefont{V.~B.} \bibnamefont{{Adya}}},
  \bibnamefont{et~al.}, \bibinfo{journal}{\nat} \textbf{\bibinfo{volume}{551}},
  \bibinfo{pages}{85} (\bibinfo{year}{2017}{\natexlab{a}}),
  \eprint{1710.05835}.

\bibitem[{\citenamefont{{Nissanke} et~al.}(2013)\citenamefont{{Nissanke},
  {Holz}, {Dalal}, {Hughes}, {Sievers}, and {Hirata}}}]{2013arXiv1307.2638N}
\bibinfo{author}{\bibfnamefont{S.}~\bibnamefont{{Nissanke}}},
  \bibinfo{author}{\bibfnamefont{D.~E.} \bibnamefont{{Holz}}},
  \bibinfo{author}{\bibfnamefont{N.}~\bibnamefont{{Dalal}}},
  \bibinfo{author}{\bibfnamefont{S.~A.} \bibnamefont{{Hughes}}},
  \bibinfo{author}{\bibfnamefont{J.~L.} \bibnamefont{{Sievers}}},
  \bibnamefont{and} \bibinfo{author}{\bibfnamefont{C.~M.}
  \bibnamefont{{Hirata}}}, \bibinfo{journal}{ArXiv e-prints}
  (\bibinfo{year}{2013}), \eprint{1307.2638}.

\bibitem[{\citenamefont{{Abbott}
  et~al.}(2016{\natexlab{a}})\citenamefont{{Abbott}, {Abbott}, {Abbott},
  {Abernathy}, {Acernese}, {Ackley}, {Adams}, {Adams}, {Addesso}, {Adhikari}
  et~al.}}]{2016LRR....19....1A}
\bibinfo{author}{\bibfnamefont{B.~P.} \bibnamefont{{Abbott}}},
  \bibinfo{author}{\bibfnamefont{R.}~\bibnamefont{{Abbott}}},
  \bibinfo{author}{\bibfnamefont{T.~D.} \bibnamefont{{Abbott}}},
  \bibinfo{author}{\bibfnamefont{M.~R.} \bibnamefont{{Abernathy}}},
  \bibinfo{author}{\bibfnamefont{F.}~\bibnamefont{{Acernese}}},
  \bibinfo{author}{\bibfnamefont{K.}~\bibnamefont{{Ackley}}},
  \bibinfo{author}{\bibfnamefont{C.}~\bibnamefont{{Adams}}},
  \bibinfo{author}{\bibfnamefont{T.}~\bibnamefont{{Adams}}},
  \bibinfo{author}{\bibfnamefont{P.}~\bibnamefont{{Addesso}}},
  \bibinfo{author}{\bibfnamefont{R.~X.} \bibnamefont{{Adhikari}}},
  \bibnamefont{et~al.}, \bibinfo{journal}{Living Reviews in Relativity}
  \textbf{\bibinfo{volume}{19}}, \bibinfo{eid}{1}
  (\bibinfo{year}{2016}{\natexlab{a}}), \eprint{1304.0670}.

\bibitem[{\citenamefont{{Chen} et~al.}(2017{\natexlab{a}})\citenamefont{{Chen},
  {Fishbach}, and {Holz}}}]{2017arXiv171206531C}
\bibinfo{author}{\bibfnamefont{H.-Y.} \bibnamefont{{Chen}}},
  \bibinfo{author}{\bibfnamefont{M.}~\bibnamefont{{Fishbach}}},
  \bibnamefont{and} \bibinfo{author}{\bibfnamefont{D.~E.}
  \bibnamefont{{Holz}}}, \bibinfo{journal}{ArXiv e-prints}
  (\bibinfo{year}{2017}{\natexlab{a}}), \eprint{1712.06531}.

\bibitem[{\citenamefont{{Holz} and {Hughes}}(2005)}]{2005ApJ...629...15H}
\bibinfo{author}{\bibfnamefont{D.~E.} \bibnamefont{{Holz}}} \bibnamefont{and}
  \bibinfo{author}{\bibfnamefont{S.~A.} \bibnamefont{{Hughes}}},
  \bibinfo{journal}{\apj} \textbf{\bibinfo{volume}{629}}, \bibinfo{pages}{15}
  (\bibinfo{year}{2005}), \eprint{astro-ph/0504616}.

\bibitem[{\citenamefont{{Messenger} and {Read}}(2012)}]{2012PhRvL.108i1101M}
\bibinfo{author}{\bibfnamefont{C.}~\bibnamefont{{Messenger}}} \bibnamefont{and}
  \bibinfo{author}{\bibfnamefont{J.}~\bibnamefont{{Read}}},
  \bibinfo{journal}{Physical Review Letters} \textbf{\bibinfo{volume}{108}},
  \bibinfo{eid}{091101} (\bibinfo{year}{2012}), \eprint{1107.5725}.

\bibitem[{\citenamefont{{Del Pozzo}}(2012)}]{2012PhRvD..86d3011D}
\bibinfo{author}{\bibfnamefont{W.}~\bibnamefont{{Del Pozzo}}},
  \bibinfo{journal}{\prd} \textbf{\bibinfo{volume}{86}}, \bibinfo{eid}{043011}
  (\bibinfo{year}{2012}), \eprint{1108.1317}.

\bibitem[{\citenamefont{{Del Pozzo} et~al.}(2017)\citenamefont{{Del Pozzo},
  {Li}, and {Messenger}}}]{2017PhRvD..95d3502D}
\bibinfo{author}{\bibfnamefont{W.}~\bibnamefont{{Del Pozzo}}},
  \bibinfo{author}{\bibfnamefont{T.~G.~F.} \bibnamefont{{Li}}},
  \bibnamefont{and}
  \bibinfo{author}{\bibfnamefont{C.}~\bibnamefont{{Messenger}}},
  \bibinfo{journal}{\prd} \textbf{\bibinfo{volume}{95}}, \bibinfo{eid}{043502}
  (\bibinfo{year}{2017}), \eprint{1506.06590}.

\bibitem[{\citenamefont{{Taylor} et~al.}(2012)\citenamefont{{Taylor}, {Gair},
  and {Mandel}}}]{2012PhRvD..85b3535T}
\bibinfo{author}{\bibfnamefont{S.~R.} \bibnamefont{{Taylor}}},
  \bibinfo{author}{\bibfnamefont{J.~R.} \bibnamefont{{Gair}}},
  \bibnamefont{and} \bibinfo{author}{\bibfnamefont{I.}~\bibnamefont{{Mandel}}},
  \bibinfo{journal}{\prd} \textbf{\bibinfo{volume}{85}}, \bibinfo{eid}{023535}
  (\bibinfo{year}{2012}), \eprint{1108.5161}.

\bibitem[{\citenamefont{{Abbott}
  et~al.}(2016{\natexlab{b}})\citenamefont{{Abbott}, {Abbott}, {Abbott},
  {Abernathy}, {Acernese}, {Ackley}, {Adams}, {Adams}, {Addesso}, {Adhikari}
  et~al.}}]{GW150914-PARAMESTIM}
\bibinfo{author}{\bibfnamefont{B.~P.} \bibnamefont{{Abbott}}},
  \bibinfo{author}{\bibfnamefont{R.}~\bibnamefont{{Abbott}}},
  \bibinfo{author}{\bibfnamefont{T.~D.} \bibnamefont{{Abbott}}},
  \bibinfo{author}{\bibfnamefont{M.~R.} \bibnamefont{{Abernathy}}},
  \bibinfo{author}{\bibfnamefont{F.}~\bibnamefont{{Acernese}}},
  \bibinfo{author}{\bibfnamefont{K.}~\bibnamefont{{Ackley}}},
  \bibinfo{author}{\bibfnamefont{C.}~\bibnamefont{{Adams}}},
  \bibinfo{author}{\bibfnamefont{T.}~\bibnamefont{{Adams}}},
  \bibinfo{author}{\bibfnamefont{P.}~\bibnamefont{{Addesso}}},
  \bibinfo{author}{\bibfnamefont{R.~X.} \bibnamefont{{Adhikari}}},
  \bibnamefont{et~al.}, \bibinfo{journal}{Physical Review Letters}
  \textbf{\bibinfo{volume}{116}}, \bibinfo{eid}{241102}
  (\bibinfo{year}{2016}{\natexlab{b}}), \eprint{1602.03840}.

\bibitem[{\citenamefont{{Vitale} and {Zanolin}}(2011)}]{2011PhRvD..84j4020V}
\bibinfo{author}{\bibfnamefont{S.}~\bibnamefont{{Vitale}}} \bibnamefont{and}
  \bibinfo{author}{\bibfnamefont{M.}~\bibnamefont{{Zanolin}}},
  \bibinfo{journal}{\prd} \textbf{\bibinfo{volume}{84}}, \bibinfo{eid}{104020}
  (\bibinfo{year}{2011}), \eprint{1108.2410}.

\bibitem[{\citenamefont{{Rodriguez} et~al.}(2014)\citenamefont{{Rodriguez},
  {Farr}, {Raymond}, {Farr}, {Littenberg}, {Fazi}, and
  {Kalogera}}}]{2014ApJ...784..119R}
\bibinfo{author}{\bibfnamefont{C.~L.} \bibnamefont{{Rodriguez}}},
  \bibinfo{author}{\bibfnamefont{B.}~\bibnamefont{{Farr}}},
  \bibinfo{author}{\bibfnamefont{V.}~\bibnamefont{{Raymond}}},
  \bibinfo{author}{\bibfnamefont{W.~M.} \bibnamefont{{Farr}}},
  \bibinfo{author}{\bibfnamefont{T.~B.} \bibnamefont{{Littenberg}}},
  \bibinfo{author}{\bibfnamefont{D.}~\bibnamefont{{Fazi}}}, \bibnamefont{and}
  \bibinfo{author}{\bibfnamefont{V.}~\bibnamefont{{Kalogera}}},
  \bibinfo{journal}{\apj} \textbf{\bibinfo{volume}{784}}, \bibinfo{eid}{119}
  (\bibinfo{year}{2014}), \eprint{1309.3273}.

\bibitem[{\citenamefont{{Vitale} and {Del Pozzo}}(2014)}]{2014PhRvD..89b2002V}
\bibinfo{author}{\bibfnamefont{S.}~\bibnamefont{{Vitale}}} \bibnamefont{and}
  \bibinfo{author}{\bibfnamefont{W.}~\bibnamefont{{Del Pozzo}}},
  \bibinfo{journal}{\prd} \textbf{\bibinfo{volume}{89}}, \bibinfo{eid}{022002}
  (\bibinfo{year}{2014}), \eprint{1311.2057}.

\bibitem[{\citenamefont{{Farr} et~al.}(2016)\citenamefont{{Farr}, {Berry},
  {Farr}, {Haster}, {Middleton}, {Cannon}, {Graff}, {Hanna}, {Mandel}, {Pankow}
  et~al.}}]{2016ApJ...825..116F}
\bibinfo{author}{\bibfnamefont{B.}~\bibnamefont{{Farr}}},
  \bibinfo{author}{\bibfnamefont{C.~P.~L.} \bibnamefont{{Berry}}},
  \bibinfo{author}{\bibfnamefont{W.~M.} \bibnamefont{{Farr}}},
  \bibinfo{author}{\bibfnamefont{C.-J.} \bibnamefont{{Haster}}},
  \bibinfo{author}{\bibfnamefont{H.}~\bibnamefont{{Middleton}}},
  \bibinfo{author}{\bibfnamefont{K.}~\bibnamefont{{Cannon}}},
  \bibinfo{author}{\bibfnamefont{P.~B.} \bibnamefont{{Graff}}},
  \bibinfo{author}{\bibfnamefont{C.}~\bibnamefont{{Hanna}}},
  \bibinfo{author}{\bibfnamefont{I.}~\bibnamefont{{Mandel}}},
  \bibinfo{author}{\bibfnamefont{C.}~\bibnamefont{{Pankow}}},
  \bibnamefont{et~al.}, \bibinfo{journal}{\apj} \textbf{\bibinfo{volume}{825}},
  \bibinfo{eid}{116} (\bibinfo{year}{2016}), \eprint{1508.05336}.

\bibitem[{\citenamefont{{Paczynski}}(1991)}]{1991AcA....41..257P}
\bibinfo{author}{\bibfnamefont{B.}~\bibnamefont{{Paczynski}}},
  \bibinfo{journal}{Acta Astronomica} \textbf{\bibinfo{volume}{41}},
  \bibinfo{pages}{257} (\bibinfo{year}{1991}).

\bibitem[{\citenamefont{{Narayan} et~al.}(1992)\citenamefont{{Narayan},
  {Paczynski}, and {Piran}}}]{1992ApJ...395L..83N}
\bibinfo{author}{\bibfnamefont{R.}~\bibnamefont{{Narayan}}},
  \bibinfo{author}{\bibfnamefont{B.}~\bibnamefont{{Paczynski}}},
  \bibnamefont{and} \bibinfo{author}{\bibfnamefont{T.}~\bibnamefont{{Piran}}},
  \bibinfo{journal}{\apjl} \textbf{\bibinfo{volume}{395}}, \bibinfo{pages}{L83}
  (\bibinfo{year}{1992}), \eprint{astro-ph/9204001}.

\bibitem[{\citenamefont{{Janka} et~al.}(1999)\citenamefont{{Janka}, {Eberl},
  {Ruffert}, and {Fryer}}}]{1999ApJ...527L..39J}
\bibinfo{author}{\bibfnamefont{H.-T.} \bibnamefont{{Janka}}},
  \bibinfo{author}{\bibfnamefont{T.}~\bibnamefont{{Eberl}}},
  \bibinfo{author}{\bibfnamefont{M.}~\bibnamefont{{Ruffert}}},
  \bibnamefont{and} \bibinfo{author}{\bibfnamefont{C.~L.}
  \bibnamefont{{Fryer}}}, \bibinfo{journal}{\apjl}
  \textbf{\bibinfo{volume}{527}}, \bibinfo{pages}{L39} (\bibinfo{year}{1999}),
  \eprint{astro-ph/9908290}.

\bibitem[{\citenamefont{{Deaton} et~al.}(2013)\citenamefont{{Deaton}, {Duez},
  {Foucart}, {O'Connor}, {Ott}, {Kidder}, {Muhlberger}, {Scheel}, and
  {Szilagyi}}}]{2013ApJ...776...47D}
\bibinfo{author}{\bibfnamefont{M.~B.} \bibnamefont{{Deaton}}},
  \bibinfo{author}{\bibfnamefont{M.~D.} \bibnamefont{{Duez}}},
  \bibinfo{author}{\bibfnamefont{F.}~\bibnamefont{{Foucart}}},
  \bibinfo{author}{\bibfnamefont{E.}~\bibnamefont{{O'Connor}}},
  \bibinfo{author}{\bibfnamefont{C.~D.} \bibnamefont{{Ott}}},
  \bibinfo{author}{\bibfnamefont{L.~E.} \bibnamefont{{Kidder}}},
  \bibinfo{author}{\bibfnamefont{C.~D.} \bibnamefont{{Muhlberger}}},
  \bibinfo{author}{\bibfnamefont{M.~A.} \bibnamefont{{Scheel}}},
  \bibnamefont{and}
  \bibinfo{author}{\bibfnamefont{B.}~\bibnamefont{{Szilagyi}}},
  \bibinfo{journal}{\apj} \textbf{\bibinfo{volume}{776}}, \bibinfo{eid}{47}
  (\bibinfo{year}{2013}), \eprint{1304.3384}.

\bibitem[{\citenamefont{{Kyutoku} et~al.}(2018)\citenamefont{{Kyutoku},
  {Kiuchi}, {Sekiguchi}, {Shibata}, and {Taniguchi}}}]{2018PhRvD..97b3009K}
\bibinfo{author}{\bibfnamefont{K.}~\bibnamefont{{Kyutoku}}},
  \bibinfo{author}{\bibfnamefont{K.}~\bibnamefont{{Kiuchi}}},
  \bibinfo{author}{\bibfnamefont{Y.}~\bibnamefont{{Sekiguchi}}},
  \bibinfo{author}{\bibfnamefont{M.}~\bibnamefont{{Shibata}}},
  \bibnamefont{and}
  \bibinfo{author}{\bibfnamefont{K.}~\bibnamefont{{Taniguchi}}},
  \bibinfo{journal}{\prd} \textbf{\bibinfo{volume}{97}}, \bibinfo{eid}{023009}
  (\bibinfo{year}{2018}), \eprint{1710.00827}.

\bibitem[{\citenamefont{{Etienne} et~al.}(2009)\citenamefont{{Etienne}, {Liu},
  {Shapiro}, and {Baumgarte}}}]{2009PhRvD..79d4024E}
\bibinfo{author}{\bibfnamefont{Z.~B.} \bibnamefont{{Etienne}}},
  \bibinfo{author}{\bibfnamefont{Y.~T.} \bibnamefont{{Liu}}},
  \bibinfo{author}{\bibfnamefont{S.~L.} \bibnamefont{{Shapiro}}},
  \bibnamefont{and} \bibinfo{author}{\bibfnamefont{T.~W.}
  \bibnamefont{{Baumgarte}}}, \bibinfo{journal}{\prd}
  \textbf{\bibinfo{volume}{79}}, \bibinfo{eid}{044024} (\bibinfo{year}{2009}),
  \eprint{0812.2245}.

\bibitem[{\citenamefont{{Foucart} et~al.}(2013)\citenamefont{{Foucart},
  {Deaton}, {Duez}, {Kidder}, {MacDonald}, {Ott}, {Pfeiffer}, {Scheel},
  {Szilagyi}, and {Teukolsky}}}]{2013PhRvD..87h4006F}
\bibinfo{author}{\bibfnamefont{F.}~\bibnamefont{{Foucart}}},
  \bibinfo{author}{\bibfnamefont{M.~B.} \bibnamefont{{Deaton}}},
  \bibinfo{author}{\bibfnamefont{M.~D.} \bibnamefont{{Duez}}},
  \bibinfo{author}{\bibfnamefont{L.~E.} \bibnamefont{{Kidder}}},
  \bibinfo{author}{\bibfnamefont{I.}~\bibnamefont{{MacDonald}}},
  \bibinfo{author}{\bibfnamefont{C.~D.} \bibnamefont{{Ott}}},
  \bibinfo{author}{\bibfnamefont{H.~P.} \bibnamefont{{Pfeiffer}}},
  \bibinfo{author}{\bibfnamefont{M.~A.} \bibnamefont{{Scheel}}},
  \bibinfo{author}{\bibfnamefont{B.}~\bibnamefont{{Szilagyi}}},
  \bibnamefont{and} \bibinfo{author}{\bibfnamefont{S.~A.}
  \bibnamefont{{Teukolsky}}}, \bibinfo{journal}{\prd}
  \textbf{\bibinfo{volume}{87}}, \bibinfo{eid}{084006} (\bibinfo{year}{2013}),
  \eprint{1212.4810}.

\bibitem[{\citenamefont{{Kyutoku} et~al.}(2011)\citenamefont{{Kyutoku},
  {Okawa}, {Shibata}, and {Taniguchi}}}]{2011PhRvD..84f4018K}
\bibinfo{author}{\bibfnamefont{K.}~\bibnamefont{{Kyutoku}}},
  \bibinfo{author}{\bibfnamefont{H.}~\bibnamefont{{Okawa}}},
  \bibinfo{author}{\bibfnamefont{M.}~\bibnamefont{{Shibata}}},
  \bibnamefont{and}
  \bibinfo{author}{\bibfnamefont{K.}~\bibnamefont{{Taniguchi}}},
  \bibinfo{journal}{\prd} \textbf{\bibinfo{volume}{84}}, \bibinfo{eid}{064018}
  (\bibinfo{year}{2011}), \eprint{1108.1189}.

\bibitem[{\citenamefont{{Foucart}}(2012)}]{2012PhRvD..86l4007F}
\bibinfo{author}{\bibfnamefont{F.}~\bibnamefont{{Foucart}}},
  \bibinfo{journal}{\prd} \textbf{\bibinfo{volume}{86}}, \bibinfo{eid}{124007}
  (\bibinfo{year}{2012}), \eprint{1207.6304}.

\bibitem[{\citenamefont{{Lovelace} et~al.}(2013)\citenamefont{{Lovelace},
  {Duez}, {Foucart}, {Kidder}, {Pfeiffer}, {Scheel}, and
  {Szil{\'a}gyi}}}]{2013CQGra..30m5004L}
\bibinfo{author}{\bibfnamefont{G.}~\bibnamefont{{Lovelace}}},
  \bibinfo{author}{\bibfnamefont{M.~D.} \bibnamefont{{Duez}}},
  \bibinfo{author}{\bibfnamefont{F.}~\bibnamefont{{Foucart}}},
  \bibinfo{author}{\bibfnamefont{L.~E.} \bibnamefont{{Kidder}}},
  \bibinfo{author}{\bibfnamefont{H.~P.} \bibnamefont{{Pfeiffer}}},
  \bibinfo{author}{\bibfnamefont{M.~A.} \bibnamefont{{Scheel}}},
  \bibnamefont{and}
  \bibinfo{author}{\bibfnamefont{B.}~\bibnamefont{{Szil{\'a}gyi}}},
  \bibinfo{journal}{Classical and Quantum Gravity}
  \textbf{\bibinfo{volume}{30}}, \bibinfo{eid}{135004} (\bibinfo{year}{2013}),
  \eprint{1302.6297}.

\bibitem[{\citenamefont{{Just} et~al.}(2015)\citenamefont{{Just}, {Bauswein},
  {Pulpillo}, {Goriely}, and {Janka}}}]{2015MNRAS.448..541J}
\bibinfo{author}{\bibfnamefont{O.}~\bibnamefont{{Just}}},
  \bibinfo{author}{\bibfnamefont{A.}~\bibnamefont{{Bauswein}}},
  \bibinfo{author}{\bibfnamefont{R.~A.} \bibnamefont{{Pulpillo}}},
  \bibinfo{author}{\bibfnamefont{S.}~\bibnamefont{{Goriely}}},
  \bibnamefont{and} \bibinfo{author}{\bibfnamefont{H.-T.}
  \bibnamefont{{Janka}}}, \bibinfo{journal}{\mnras}
  \textbf{\bibinfo{volume}{448}}, \bibinfo{pages}{541} (\bibinfo{year}{2015}),
  \eprint{1406.2687}.

\bibitem[{\citenamefont{{Fern{\'a}ndez}
  et~al.}(2015)\citenamefont{{Fern{\'a}ndez}, {Kasen}, {Metzger}, and
  {Quataert}}}]{2015MNRAS.446..750F}
\bibinfo{author}{\bibfnamefont{R.}~\bibnamefont{{Fern{\'a}ndez}}},
  \bibinfo{author}{\bibfnamefont{D.}~\bibnamefont{{Kasen}}},
  \bibinfo{author}{\bibfnamefont{B.~D.} \bibnamefont{{Metzger}}},
  \bibnamefont{and}
  \bibinfo{author}{\bibfnamefont{E.}~\bibnamefont{{Quataert}}},
  \bibinfo{journal}{\mnras} \textbf{\bibinfo{volume}{446}},
  \bibinfo{pages}{750} (\bibinfo{year}{2015}), \eprint{1409.4426}.

\bibitem[{\citenamefont{{Vecchio}}(2004)}]{2004PhRvD..70d2001V}
\bibinfo{author}{\bibfnamefont{A.}~\bibnamefont{{Vecchio}}},
  \bibinfo{journal}{\prd} \textbf{\bibinfo{volume}{70}}, \bibinfo{eid}{042001}
  (\bibinfo{year}{2004}), \eprint{astro-ph/0304051}.

\bibitem[{\citenamefont{{Vitale} et~al.}(2014)\citenamefont{{Vitale}, {Lynch},
  {Veitch}, {Raymond}, and {Sturani}}}]{2014PhRvL.112y1101V}
\bibinfo{author}{\bibfnamefont{S.}~\bibnamefont{{Vitale}}},
  \bibinfo{author}{\bibfnamefont{R.}~\bibnamefont{{Lynch}}},
  \bibinfo{author}{\bibfnamefont{J.}~\bibnamefont{{Veitch}}},
  \bibinfo{author}{\bibfnamefont{V.}~\bibnamefont{{Raymond}}},
  \bibnamefont{and}
  \bibinfo{author}{\bibfnamefont{R.}~\bibnamefont{{Sturani}}},
  \bibinfo{journal}{Physical Review Letters} \textbf{\bibinfo{volume}{112}},
  \bibinfo{eid}{251101} (\bibinfo{year}{2014}), \eprint{1403.0129}.

\bibitem[{\citenamefont{{McWilliams} et~al.}(2011)\citenamefont{{McWilliams},
  {Lang}, {Baker}, and {Thorpe}}}]{2011PhRvD..84f4003M}
\bibinfo{author}{\bibfnamefont{S.~T.} \bibnamefont{{McWilliams}}},
  \bibinfo{author}{\bibfnamefont{R.~N.} \bibnamefont{{Lang}}},
  \bibinfo{author}{\bibfnamefont{J.~G.} \bibnamefont{{Baker}}},
  \bibnamefont{and} \bibinfo{author}{\bibfnamefont{J.~I.}
  \bibnamefont{{Thorpe}}}, \bibinfo{journal}{\prd}
  \textbf{\bibinfo{volume}{84}}, \bibinfo{eid}{064003} (\bibinfo{year}{2011}),
  \eprint{1104.5650}.

\bibitem[{\citenamefont{{Klein} et~al.}(2016)\citenamefont{{Klein}, {Barausse},
  {Sesana}, {Petiteau}, {Berti}, {Babak}, {Gair}, {Aoudia}, {Hinder}, {Ohme}
  et~al.}}]{2016PhRvD..93b4003K}
\bibinfo{author}{\bibfnamefont{A.}~\bibnamefont{{Klein}}},
  \bibinfo{author}{\bibfnamefont{E.}~\bibnamefont{{Barausse}}},
  \bibinfo{author}{\bibfnamefont{A.}~\bibnamefont{{Sesana}}},
  \bibinfo{author}{\bibfnamefont{A.}~\bibnamefont{{Petiteau}}},
  \bibinfo{author}{\bibfnamefont{E.}~\bibnamefont{{Berti}}},
  \bibinfo{author}{\bibfnamefont{S.}~\bibnamefont{{Babak}}},
  \bibinfo{author}{\bibfnamefont{J.}~\bibnamefont{{Gair}}},
  \bibinfo{author}{\bibfnamefont{S.}~\bibnamefont{{Aoudia}}},
  \bibinfo{author}{\bibfnamefont{I.}~\bibnamefont{{Hinder}}},
  \bibinfo{author}{\bibfnamefont{F.}~\bibnamefont{{Ohme}}},
  \bibnamefont{et~al.}, \bibinfo{journal}{\prd} \textbf{\bibinfo{volume}{93}},
  \bibinfo{eid}{024003} (\bibinfo{year}{2016}), \eprint{1511.05581}.

\bibitem[{\citenamefont{Apostolatos et~al.}(1994)\citenamefont{Apostolatos,
  Cutler, Sussman, and Thorne}}]{PhysRevD.49.6274}
\bibinfo{author}{\bibfnamefont{T.~A.} \bibnamefont{Apostolatos}},
  \bibinfo{author}{\bibfnamefont{C.}~\bibnamefont{Cutler}},
  \bibinfo{author}{\bibfnamefont{G.~J.} \bibnamefont{Sussman}},
  \bibnamefont{and} \bibinfo{author}{\bibfnamefont{K.~S.}
  \bibnamefont{Thorne}}, \bibinfo{journal}{Phys. Rev. D}
  \textbf{\bibinfo{volume}{49}}, \bibinfo{pages}{6274} (\bibinfo{year}{1994}),
  \urlprefix\url{http://link.aps.org/doi/10.1103/PhysRevD.49.6274}.

\bibitem[{\citenamefont{Vitale et~al.}(2017)\citenamefont{Vitale, Lynch,
  Raymond, Sturani, Veitch, and Graff}}]{Vitale:2016avz}
\bibinfo{author}{\bibfnamefont{S.}~\bibnamefont{Vitale}},
  \bibinfo{author}{\bibfnamefont{R.}~\bibnamefont{Lynch}},
  \bibinfo{author}{\bibfnamefont{V.}~\bibnamefont{Raymond}},
  \bibinfo{author}{\bibfnamefont{R.}~\bibnamefont{Sturani}},
  \bibinfo{author}{\bibfnamefont{J.}~\bibnamefont{Veitch}}, \bibnamefont{and}
  \bibinfo{author}{\bibfnamefont{P.}~\bibnamefont{Graff}},
  \bibinfo{journal}{Phys. Rev. D} \textbf{\bibinfo{volume}{95}},
  \bibinfo{pages}{064053} (\bibinfo{year}{2017}), \eprint{1611.01122}.

\bibitem[{\citenamefont{{Abbott}
  et~al.}(2016{\natexlab{c}})\citenamefont{{Abbott}, {Abbott}, {Abbott},
  {Abernathy}, {Acernese}, {Ackley}, {Adams}, {Adams}, {Addesso}, {Adhikari}
  et~al.}}]{2016ApJ...832L..21A}
\bibinfo{author}{\bibfnamefont{B.~P.} \bibnamefont{{Abbott}}},
  \bibinfo{author}{\bibfnamefont{R.}~\bibnamefont{{Abbott}}},
  \bibinfo{author}{\bibfnamefont{T.~D.} \bibnamefont{{Abbott}}},
  \bibinfo{author}{\bibfnamefont{M.~R.} \bibnamefont{{Abernathy}}},
  \bibinfo{author}{\bibfnamefont{F.}~\bibnamefont{{Acernese}}},
  \bibinfo{author}{\bibfnamefont{K.}~\bibnamefont{{Ackley}}},
  \bibinfo{author}{\bibfnamefont{C.}~\bibnamefont{{Adams}}},
  \bibinfo{author}{\bibfnamefont{T.}~\bibnamefont{{Adams}}},
  \bibinfo{author}{\bibfnamefont{P.}~\bibnamefont{{Addesso}}},
  \bibinfo{author}{\bibfnamefont{R.~X.} \bibnamefont{{Adhikari}}},
  \bibnamefont{et~al.}, \bibinfo{journal}{\apjl}
  \textbf{\bibinfo{volume}{832}}, \bibinfo{eid}{L21}
  (\bibinfo{year}{2016}{\natexlab{c}}), \eprint{1607.07456}.

\bibitem[{\citenamefont{{Vallisneri}}(2008)}]{2008PhRvD..77d2001V}
\bibinfo{author}{\bibfnamefont{M.}~\bibnamefont{{Vallisneri}}},
  \bibinfo{journal}{\prd} \textbf{\bibinfo{volume}{77}}, \bibinfo{eid}{042001}
  (\bibinfo{year}{2008}), \eprint{gr-qc/0703086}.

\bibitem[{\citenamefont{Abbott et~al.}(2013)}]{Aasi:2013wya}
\bibinfo{author}{\bibfnamefont{B.~P.} \bibnamefont{Abbott}}
  \bibnamefont{et~al.} (\bibinfo{collaboration}{VIRGO, LIGO Scientific})
  (\bibinfo{year}{2013}), \bibinfo{note}{[Living Rev. Rel.19,1(2016)]},
  \eprint{1304.0670}.

\bibitem[{\citenamefont{{Zhao} et~al.}(2017)\citenamefont{{Zhao}, {Kesden}, and
  {Gerosa}}}]{2017PhRvD..96b4007Z}
\bibinfo{author}{\bibfnamefont{X.}~\bibnamefont{{Zhao}}},
  \bibinfo{author}{\bibfnamefont{M.}~\bibnamefont{{Kesden}}}, \bibnamefont{and}
  \bibinfo{author}{\bibfnamefont{D.}~\bibnamefont{{Gerosa}}},
  \bibinfo{journal}{\prd} \textbf{\bibinfo{volume}{96}}, \bibinfo{eid}{024007}
  (\bibinfo{year}{2017}), \eprint{1705.02369}.

\bibitem[{\citenamefont{{Schmidt} et~al.}(2015)\citenamefont{{Schmidt}, {Ohme},
  and {Hannam}}}]{2015PhRvD..91b4043S}
\bibinfo{author}{\bibfnamefont{P.}~\bibnamefont{{Schmidt}}},
  \bibinfo{author}{\bibfnamefont{F.}~\bibnamefont{{Ohme}}}, \bibnamefont{and}
  \bibinfo{author}{\bibfnamefont{M.}~\bibnamefont{{Hannam}}},
  \bibinfo{journal}{\prd} \textbf{\bibinfo{volume}{91}}, \bibinfo{eid}{024043}
  (\bibinfo{year}{2015}), \eprint{1408.1810}.

\bibitem[{\citenamefont{Hannam et~al.}(2014)\citenamefont{Hannam, Schmidt,
  Boh{\'e}, Haegel, Husa, Ohme, Pratten, and P{\"u}rrer}}]{Hannam:2013oca}
\bibinfo{author}{\bibfnamefont{M.}~\bibnamefont{Hannam}},
  \bibinfo{author}{\bibfnamefont{P.}~\bibnamefont{Schmidt}},
  \bibinfo{author}{\bibfnamefont{A.}~\bibnamefont{Boh{\'e}}},
  \bibinfo{author}{\bibfnamefont{L.}~\bibnamefont{Haegel}},
  \bibinfo{author}{\bibfnamefont{S.}~\bibnamefont{Husa}},
  \bibinfo{author}{\bibfnamefont{F.}~\bibnamefont{Ohme}},
  \bibinfo{author}{\bibfnamefont{G.}~\bibnamefont{Pratten}}, \bibnamefont{and}
  \bibinfo{author}{\bibfnamefont{M.}~\bibnamefont{P{\"u}rrer}},
  \bibinfo{journal}{Phys. Rev. Lett.} \textbf{\bibinfo{volume}{113}},
  \bibinfo{pages}{151101} (\bibinfo{year}{2014}), \eprint{1308.3271}.

\bibitem[{\citenamefont{{Abbott}
  et~al.}(2017{\natexlab{b}})\citenamefont{{Abbott}, {Abbott}, {Abbott},
  {Acernese}, {Ackley}, {Adams}, {Adams}, {Addesso}, {Adhikari}, {Adya}
  et~al.}}]{2017PhRvL.119p1101A}
\bibinfo{author}{\bibfnamefont{B.~P.} \bibnamefont{{Abbott}}},
  \bibinfo{author}{\bibfnamefont{R.}~\bibnamefont{{Abbott}}},
  \bibinfo{author}{\bibfnamefont{T.~D.} \bibnamefont{{Abbott}}},
  \bibinfo{author}{\bibfnamefont{F.}~\bibnamefont{{Acernese}}},
  \bibinfo{author}{\bibfnamefont{K.}~\bibnamefont{{Ackley}}},
  \bibinfo{author}{\bibfnamefont{C.}~\bibnamefont{{Adams}}},
  \bibinfo{author}{\bibfnamefont{T.}~\bibnamefont{{Adams}}},
  \bibinfo{author}{\bibfnamefont{P.}~\bibnamefont{{Addesso}}},
  \bibinfo{author}{\bibfnamefont{R.~X.} \bibnamefont{{Adhikari}}},
  \bibinfo{author}{\bibfnamefont{V.~B.} \bibnamefont{{Adya}}},
  \bibnamefont{et~al.}, \bibinfo{journal}{Physical Review Letters}
  \textbf{\bibinfo{volume}{119}}, \bibinfo{eid}{161101}
  (\bibinfo{year}{2017}{\natexlab{b}}), \eprint{1710.05832}.

\bibitem[{\citenamefont{{Chen} et~al.}(2017{\natexlab{b}})\citenamefont{{Chen},
  {Essick}, {Vitale}, {Holz}, and {Katsavounidis}}}]{2017ApJ...835...31C}
\bibinfo{author}{\bibfnamefont{H.-Y.} \bibnamefont{{Chen}}},
  \bibinfo{author}{\bibfnamefont{R.}~\bibnamefont{{Essick}}},
  \bibinfo{author}{\bibfnamefont{S.}~\bibnamefont{{Vitale}}},
  \bibinfo{author}{\bibfnamefont{D.~E.} \bibnamefont{{Holz}}},
  \bibnamefont{and}
  \bibinfo{author}{\bibfnamefont{E.}~\bibnamefont{{Katsavounidis}}},
  \bibinfo{journal}{\apj} \textbf{\bibinfo{volume}{835}}, \bibinfo{eid}{31}
  (\bibinfo{year}{2017}{\natexlab{b}}), \eprint{1608.00164}.

\bibitem[{\citenamefont{{Veitch} et~al.}(2015)\citenamefont{{Veitch},
  {Raymond}, {Farr}, {Farr}, {Graff}, {Vitale}, {Aylott}, {Blackburn},
  {Christensen}, {Coughlin} et~al.}}]{2015PhRvD..91d2003V}
\bibinfo{author}{\bibfnamefont{J.}~\bibnamefont{{Veitch}}},
  \bibinfo{author}{\bibfnamefont{V.}~\bibnamefont{{Raymond}}},
  \bibinfo{author}{\bibfnamefont{B.}~\bibnamefont{{Farr}}},
  \bibinfo{author}{\bibfnamefont{W.}~\bibnamefont{{Farr}}},
  \bibinfo{author}{\bibfnamefont{P.}~\bibnamefont{{Graff}}},
  \bibinfo{author}{\bibfnamefont{S.}~\bibnamefont{{Vitale}}},
  \bibinfo{author}{\bibfnamefont{B.}~\bibnamefont{{Aylott}}},
  \bibinfo{author}{\bibfnamefont{K.}~\bibnamefont{{Blackburn}}},
  \bibinfo{author}{\bibfnamefont{N.}~\bibnamefont{{Christensen}}},
  \bibinfo{author}{\bibfnamefont{M.}~\bibnamefont{{Coughlin}}},
  \bibnamefont{et~al.}, \bibinfo{journal}{\prd} \textbf{\bibinfo{volume}{91}},
  \bibinfo{eid}{042003} (\bibinfo{year}{2015}), \eprint{1409.7215}.

\bibitem[{\citenamefont{{Canizares} et~al.}(2015)\citenamefont{{Canizares},
  {Field}, {Gair}, {Raymond}, {Smith}, and {Tiglio}}}]{2015PhRvL.114g1104C}
\bibinfo{author}{\bibfnamefont{P.}~\bibnamefont{{Canizares}}},
  \bibinfo{author}{\bibfnamefont{S.~E.} \bibnamefont{{Field}}},
  \bibinfo{author}{\bibfnamefont{J.}~\bibnamefont{{Gair}}},
  \bibinfo{author}{\bibfnamefont{V.}~\bibnamefont{{Raymond}}},
  \bibinfo{author}{\bibfnamefont{R.}~\bibnamefont{{Smith}}}, \bibnamefont{and}
  \bibinfo{author}{\bibfnamefont{M.}~\bibnamefont{{Tiglio}}},
  \bibinfo{journal}{Physical Review Letters} \textbf{\bibinfo{volume}{114}},
  \bibinfo{eid}{071104} (\bibinfo{year}{2015}), \eprint{1404.6284}.

\bibitem[{\citenamefont{Smith et~al.}(2016)\citenamefont{Smith, Field,
  Blackburn, Haster, P�rrer, Raymond, and Schmidt}}]{Smith:2016qas}
\bibinfo{author}{\bibfnamefont{R.}~\bibnamefont{Smith}},
  \bibinfo{author}{\bibfnamefont{S.~E.} \bibnamefont{Field}},
  \bibinfo{author}{\bibfnamefont{K.}~\bibnamefont{Blackburn}},
  \bibinfo{author}{\bibfnamefont{C.-J.} \bibnamefont{Haster}},
  \bibinfo{author}{\bibfnamefont{M.}~\bibnamefont{P�rrer}},
  \bibinfo{author}{\bibfnamefont{V.}~\bibnamefont{Raymond}}, \bibnamefont{and}
  \bibinfo{author}{\bibfnamefont{P.}~\bibnamefont{Schmidt}},
  \bibinfo{journal}{Phys. Rev.} \textbf{\bibinfo{volume}{D94}},
  \bibinfo{pages}{044031} (\bibinfo{year}{2016}), \eprint{1604.08253}.

\bibitem[{\citenamefont{{Abbott}
  et~al.}(2016{\natexlab{d}})\citenamefont{{Abbott}, {Abbott}, {Abbott},
  {Abernathy}, {Acernese}, {Ackley}, {Adams}, {Adams}, {Addesso}, {Adhikari}
  et~al.}}]{2016ApJ...818L..22A}
\bibinfo{author}{\bibfnamefont{B.~P.} \bibnamefont{{Abbott}}},
  \bibinfo{author}{\bibfnamefont{R.}~\bibnamefont{{Abbott}}},
  \bibinfo{author}{\bibfnamefont{T.~D.} \bibnamefont{{Abbott}}},
  \bibinfo{author}{\bibfnamefont{M.~R.} \bibnamefont{{Abernathy}}},
  \bibinfo{author}{\bibfnamefont{F.}~\bibnamefont{{Acernese}}},
  \bibinfo{author}{\bibfnamefont{K.}~\bibnamefont{{Ackley}}},
  \bibinfo{author}{\bibfnamefont{C.}~\bibnamefont{{Adams}}},
  \bibinfo{author}{\bibfnamefont{T.}~\bibnamefont{{Adams}}},
  \bibinfo{author}{\bibfnamefont{P.}~\bibnamefont{{Addesso}}},
  \bibinfo{author}{\bibfnamefont{R.~X.} \bibnamefont{{Adhikari}}},
  \bibnamefont{et~al.}, \bibinfo{journal}{\apjl}
  \textbf{\bibinfo{volume}{818}}, \bibinfo{eid}{L22}
  (\bibinfo{year}{2016}{\natexlab{d}}), \eprint{1602.03846}.

\bibitem[{\citenamefont{{Mooley} et~al.}(2018)\citenamefont{{Mooley}, {Nakar},
  {Hotokezaka}, {Hallinan}, {Corsi}, {Frail}, {Horesh}, {Murphy}, {Lenc},
  {Kaplan} et~al.}}]{2018Natur.554..207M}
\bibinfo{author}{\bibfnamefont{K.~P.} \bibnamefont{{Mooley}}},
  \bibinfo{author}{\bibfnamefont{E.}~\bibnamefont{{Nakar}}},
  \bibinfo{author}{\bibfnamefont{K.}~\bibnamefont{{Hotokezaka}}},
  \bibinfo{author}{\bibfnamefont{G.}~\bibnamefont{{Hallinan}}},
  \bibinfo{author}{\bibfnamefont{A.}~\bibnamefont{{Corsi}}},
  \bibinfo{author}{\bibfnamefont{D.~A.} \bibnamefont{{Frail}}},
  \bibinfo{author}{\bibfnamefont{A.}~\bibnamefont{{Horesh}}},
  \bibinfo{author}{\bibfnamefont{T.}~\bibnamefont{{Murphy}}},
  \bibinfo{author}{\bibfnamefont{E.}~\bibnamefont{{Lenc}}},
  \bibinfo{author}{\bibfnamefont{D.~L.} \bibnamefont{{Kaplan}}},
  \bibnamefont{et~al.}, \bibinfo{journal}{\nat} \textbf{\bibinfo{volume}{554}},
  \bibinfo{pages}{207} (\bibinfo{year}{2018}), \eprint{1711.11573}.

\bibitem[{\citenamefont{{Lazzati} et~al.}(2017)\citenamefont{{Lazzati},
  {Perna}, {Morsony}, {L{\'o}pez-C{\'a}mara}, {Cantiello}, {Ciolfi},
  {giacomazzo}, and {Workman}}}]{2017arXiv171203237L}
\bibinfo{author}{\bibfnamefont{D.}~\bibnamefont{{Lazzati}}},
  \bibinfo{author}{\bibfnamefont{R.}~\bibnamefont{{Perna}}},
  \bibinfo{author}{\bibfnamefont{B.~J.} \bibnamefont{{Morsony}}},
  \bibinfo{author}{\bibfnamefont{D.}~\bibnamefont{{L{\'o}pez-C{\'a}mara}}},
  \bibinfo{author}{\bibfnamefont{M.}~\bibnamefont{{Cantiello}}},
  \bibinfo{author}{\bibfnamefont{R.}~\bibnamefont{{Ciolfi}}},
  \bibinfo{author}{\bibfnamefont{B.}~\bibnamefont{{giacomazzo}}},
  \bibnamefont{and} \bibinfo{author}{\bibfnamefont{J.~C.}
  \bibnamefont{{Workman}}}, \bibinfo{journal}{ArXiv e-prints}
  (\bibinfo{year}{2017}), \eprint{1712.03237}.

\bibitem[{\citenamefont{{{Maggiore,~M.}}}(2007)}]{Maggiore}
\bibinfo{author}{\bibnamefont{{{Maggiore,~M.}}}},
  \emph{\bibinfo{title}{Gravitational Waves, Volume 1: Theory and Experiments}}
  (\bibinfo{publisher}{Oxford University Press}, \bibinfo{year}{2007}).

\bibitem[{\citenamefont{{Schutz}}(2011)}]{2011CQGra..28l5023S}
\bibinfo{author}{\bibfnamefont{B.~F.} \bibnamefont{{Schutz}}},
  \bibinfo{journal}{Classical and Quantum Gravity}
  \textbf{\bibinfo{volume}{28}}, \bibinfo{eid}{125023} (\bibinfo{year}{2011}),
  \eprint{1102.5421}.

\bibitem[{\citenamefont{{Chen} et~al.}(2017{\natexlab{c}})\citenamefont{{Chen},
  {Holz}, {Miller}, {Evans}, {Vitale}, and {Creighton}}}]{2017arXiv170908079C}
\bibinfo{author}{\bibfnamefont{H.-Y.} \bibnamefont{{Chen}}},
  \bibinfo{author}{\bibfnamefont{D.~E.} \bibnamefont{{Holz}}},
  \bibinfo{author}{\bibfnamefont{J.}~\bibnamefont{{Miller}}},
  \bibinfo{author}{\bibfnamefont{M.}~\bibnamefont{{Evans}}},
  \bibinfo{author}{\bibfnamefont{S.}~\bibnamefont{{Vitale}}}, \bibnamefont{and}
  \bibinfo{author}{\bibfnamefont{J.}~\bibnamefont{{Creighton}}},
  \bibinfo{journal}{ArXiv e-prints}  (\bibinfo{year}{2017}{\natexlab{c}}),
  \eprint{1709.08079}.

\bibitem[{\citenamefont{{Kyutoku} et~al.}(2015)\citenamefont{{Kyutoku}, {Ioka},
  {Okawa}, {Shibata}, and {Taniguchi}}}]{2015PhRvD..92d4028K}
\bibinfo{author}{\bibfnamefont{K.}~\bibnamefont{{Kyutoku}}},
  \bibinfo{author}{\bibfnamefont{K.}~\bibnamefont{{Ioka}}},
  \bibinfo{author}{\bibfnamefont{H.}~\bibnamefont{{Okawa}}},
  \bibinfo{author}{\bibfnamefont{M.}~\bibnamefont{{Shibata}}},
  \bibnamefont{and}
  \bibinfo{author}{\bibfnamefont{K.}~\bibnamefont{{Taniguchi}}},
  \bibinfo{journal}{\prd} \textbf{\bibinfo{volume}{92}}, \bibinfo{eid}{044028}
  (\bibinfo{year}{2015}), \eprint{1502.05402}.

\bibitem[{\citenamefont{{Ferrari} et~al.}(2010)\citenamefont{{Ferrari},
  {Gualtieri}, and {Pannarale}}}]{2010PhRvD..81f4026F}
\bibinfo{author}{\bibfnamefont{V.}~\bibnamefont{{Ferrari}}},
  \bibinfo{author}{\bibfnamefont{L.}~\bibnamefont{{Gualtieri}}},
  \bibnamefont{and}
  \bibinfo{author}{\bibfnamefont{F.}~\bibnamefont{{Pannarale}}},
  \bibinfo{journal}{\prd} \textbf{\bibinfo{volume}{81}}, \bibinfo{eid}{064026}
  (\bibinfo{year}{2010}), \eprint{0912.3692}.

\bibitem[{\citenamefont{{Ferrari} et~al.}(2009)\citenamefont{{Ferrari},
  {Gualtieri}, and {Pannarale}}}]{2009CQGra..26l5004F}
\bibinfo{author}{\bibfnamefont{V.}~\bibnamefont{{Ferrari}}},
  \bibinfo{author}{\bibfnamefont{L.}~\bibnamefont{{Gualtieri}}},
  \bibnamefont{and}
  \bibinfo{author}{\bibfnamefont{F.}~\bibnamefont{{Pannarale}}},
  \bibinfo{journal}{Classical and Quantum Gravity}
  \textbf{\bibinfo{volume}{26}}, \bibinfo{eid}{125004} (\bibinfo{year}{2009}),
  \eprint{0801.2911}.

\bibitem[{\citenamefont{Stone et~al.}(2013)\citenamefont{Stone, Loeb, and
  Berger}}]{PhysRevD.87.084053}
\bibinfo{author}{\bibfnamefont{N.}~\bibnamefont{Stone}},
  \bibinfo{author}{\bibfnamefont{A.}~\bibnamefont{Loeb}}, \bibnamefont{and}
  \bibinfo{author}{\bibfnamefont{E.}~\bibnamefont{Berger}},
  \bibinfo{journal}{Phys. Rev. D} \textbf{\bibinfo{volume}{87}},
  \bibinfo{pages}{084053} (\bibinfo{year}{2013}),
  \urlprefix\url{https://link.aps.org/doi/10.1103/PhysRevD.87.084053}.

\bibitem[{\citenamefont{{Pannarale} et~al.}(2015)\citenamefont{{Pannarale},
  {Berti}, {Kyutoku}, {Lackey}, and {Shibata}}}]{2015PhRvD..92h4050P}
\bibinfo{author}{\bibfnamefont{F.}~\bibnamefont{{Pannarale}}},
  \bibinfo{author}{\bibfnamefont{E.}~\bibnamefont{{Berti}}},
  \bibinfo{author}{\bibfnamefont{K.}~\bibnamefont{{Kyutoku}}},
  \bibinfo{author}{\bibfnamefont{B.~D.} \bibnamefont{{Lackey}}},
  \bibnamefont{and}
  \bibinfo{author}{\bibfnamefont{M.}~\bibnamefont{{Shibata}}},
  \bibinfo{journal}{\prd} \textbf{\bibinfo{volume}{92}}, \bibinfo{eid}{084050}
  (\bibinfo{year}{2015}), \eprint{1509.00512}.

\bibitem[{\citenamefont{{Liska} et~al.}(2018)\citenamefont{{Liska}, {Hesp},
  {Tchekhovskoy}, {Ingram}, {van der Klis}, and
  {Markoff}}}]{2018MNRAS.474L..81L}
\bibinfo{author}{\bibfnamefont{M.}~\bibnamefont{{Liska}}},
  \bibinfo{author}{\bibfnamefont{C.}~\bibnamefont{{Hesp}}},
  \bibinfo{author}{\bibfnamefont{A.}~\bibnamefont{{Tchekhovskoy}}},
  \bibinfo{author}{\bibfnamefont{A.}~\bibnamefont{{Ingram}}},
  \bibinfo{author}{\bibfnamefont{M.}~\bibnamefont{{van der Klis}}},
  \bibnamefont{and}
  \bibinfo{author}{\bibfnamefont{S.}~\bibnamefont{{Markoff}}},
  \bibinfo{journal}{\mnras} \textbf{\bibinfo{volume}{474}},
  \bibinfo{pages}{L81} (\bibinfo{year}{2018}), \eprint{1707.06619}.

\bibitem[{\citenamefont{{Kawaguchi} et~al.}(2016)\citenamefont{{Kawaguchi},
  {Kyutoku}, {Shibata}, and {Tanaka}}}]{2016ApJ...825...52K}
\bibinfo{author}{\bibfnamefont{K.}~\bibnamefont{{Kawaguchi}}},
  \bibinfo{author}{\bibfnamefont{K.}~\bibnamefont{{Kyutoku}}},
  \bibinfo{author}{\bibfnamefont{M.}~\bibnamefont{{Shibata}}},
  \bibnamefont{and} \bibinfo{author}{\bibfnamefont{M.}~\bibnamefont{{Tanaka}}},
  \bibinfo{journal}{\apj} \textbf{\bibinfo{volume}{825}}, \bibinfo{eid}{52}
  (\bibinfo{year}{2016}), \eprint{1601.07711}.

\bibitem[{\citenamefont{{Fairhurst}}(2011)}]{2011CQGra..28j5021F}
\bibinfo{author}{\bibfnamefont{S.}~\bibnamefont{{Fairhurst}}},
  \bibinfo{journal}{Classical and Quantum Gravity}
  \textbf{\bibinfo{volume}{28}}, \bibinfo{eid}{105021} (\bibinfo{year}{2011}),
  \eprint{1010.6192}.

\bibitem[{\citenamefont{Aso et~al.}(2013)\citenamefont{Aso, Michimura, Somiya,
  Ando, Miyakawa, Sekiguchi, Tatsumi, and Yamamoto}}]{PhysRevD.88.043007}
\bibinfo{author}{\bibfnamefont{Y.}~\bibnamefont{Aso}},
  \bibinfo{author}{\bibfnamefont{Y.}~\bibnamefont{Michimura}},
  \bibinfo{author}{\bibfnamefont{K.}~\bibnamefont{Somiya}},
  \bibinfo{author}{\bibfnamefont{M.}~\bibnamefont{Ando}},
  \bibinfo{author}{\bibfnamefont{O.}~\bibnamefont{Miyakawa}},
  \bibinfo{author}{\bibfnamefont{T.}~\bibnamefont{Sekiguchi}},
  \bibinfo{author}{\bibfnamefont{D.}~\bibnamefont{Tatsumi}}, \bibnamefont{and}
  \bibinfo{author}{\bibfnamefont{H.}~\bibnamefont{Yamamoto}}
  (\bibinfo{collaboration}{The KAGRA Collaboration}), \bibinfo{journal}{Phys.
  Rev. D} \textbf{\bibinfo{volume}{88}}, \bibinfo{pages}{043007}
  (\bibinfo{year}{2013}),
  \urlprefix\url{http://link.aps.org/doi/10.1103/PhysRevD.88.043007}.

\bibitem[{\citenamefont{{Iyer} et~al.}(2011)}]{M1100296}
\bibinfo{author}{\bibfnamefont{B.}~\bibnamefont{{Iyer}}} \bibnamefont{et~al.},
  \bibinfo{type}{Tech. Rep.} \bibinfo{number}{LIGO-M1100296}
  (\bibinfo{year}{2011}),
  \bibinfo{note}{https://dcc.ligo.org/LIGO-M1100296/public}.

\bibitem[{\citenamefont{{Vitale} et~al.}(2017)\citenamefont{{Vitale}, {Essick},
  {Katsavounidis}, {Klimenko}, and {Vedovato}}}]{2017MNRAS.466L..78V}
\bibinfo{author}{\bibfnamefont{S.}~\bibnamefont{{Vitale}}},
  \bibinfo{author}{\bibfnamefont{R.}~\bibnamefont{{Essick}}},
  \bibinfo{author}{\bibfnamefont{E.}~\bibnamefont{{Katsavounidis}}},
  \bibinfo{author}{\bibfnamefont{S.}~\bibnamefont{{Klimenko}}},
  \bibnamefont{and}
  \bibinfo{author}{\bibfnamefont{G.}~\bibnamefont{{Vedovato}}},
  \bibinfo{journal}{\mnras} \textbf{\bibinfo{volume}{466}},
  \bibinfo{pages}{L78} (\bibinfo{year}{2017}), \eprint{1611.02438}.

\bibitem[{\citenamefont{{Chen} and {Holz}}(2016)}]{2016arXiv161201471C}
\bibinfo{author}{\bibfnamefont{H.-Y.} \bibnamefont{{Chen}}} \bibnamefont{and}
  \bibinfo{author}{\bibfnamefont{D.~E.} \bibnamefont{{Holz}}},
  \bibinfo{journal}{ArXiv e-prints}  (\bibinfo{year}{2016}),
  \eprint{1612.01471}.

\bibitem[{\citenamefont{{Abbott}
  et~al.}(2017{\natexlab{c}})\citenamefont{{Abbott}, {Abbott}, {Abbott},
  {Acernese}, {Ackley}, {Adams}, {Adams}, {Addesso}, {Adhikari}, {Adya}
  et~al.}}]{2017PhRvL.119n1101A}
\bibinfo{author}{\bibfnamefont{B.~P.} \bibnamefont{{Abbott}}},
  \bibinfo{author}{\bibfnamefont{R.}~\bibnamefont{{Abbott}}},
  \bibinfo{author}{\bibfnamefont{T.~D.} \bibnamefont{{Abbott}}},
  \bibinfo{author}{\bibfnamefont{F.}~\bibnamefont{{Acernese}}},
  \bibinfo{author}{\bibfnamefont{K.}~\bibnamefont{{Ackley}}},
  \bibinfo{author}{\bibfnamefont{C.}~\bibnamefont{{Adams}}},
  \bibinfo{author}{\bibfnamefont{T.}~\bibnamefont{{Adams}}},
  \bibinfo{author}{\bibfnamefont{P.}~\bibnamefont{{Addesso}}},
  \bibinfo{author}{\bibfnamefont{R.~X.} \bibnamefont{{Adhikari}}},
  \bibinfo{author}{\bibfnamefont{V.~B.} \bibnamefont{{Adya}}},
  \bibnamefont{et~al.}, \bibinfo{journal}{Physical Review Letters}
  \textbf{\bibinfo{volume}{119}}, \bibinfo{eid}{141101}
  (\bibinfo{year}{2017}{\natexlab{c}}), \eprint{1709.09660}.

\end{thebibliography}

\end{document}